# Size and temperature dependent intraband optical studies of heavily n-doped PbS quantum dot solids

*Iñigo Ramiro[†,*], Biswajit Kundu[†], Mariona Dalmases[†], Onur Özdemir[†], María Pedrosa[†], and Gerasimos Konstantatos[†,‡,*]*

[†]ICFO—Institut de Ciències Fotòniques, The Barcelona Institute of Science and Technology, Av. Carl Friedrich Gauss, 3, 08860 Castelldefels (Barcelona), Spain

[‡]ICREA—Institució Catalana de Recerca i Estudis Avançats, Passeig Lluís Companys 23, 08010 Barcelona, Spain

**Steady-state access to intraband transitions in colloidal quantum dots (CQDs), via heavy doping, allows exploiting the electromagnetic spectrum at energies below the band gap. CQD intraband optoelectronics opens up a new path to cheap mid- and long-wavelength infrared photodetectors and light-emitting devices, which today employ mostly epitaxial materials. As a recent field of experimental research, thorough studies of the basic properties of intraband transitions in CQDs are still lacking. In this work, we investigate the size and temperature dependence of the intraband transition in heavily n-doped PbS quantum dot (QD) films. We measure the absorption coefficient of the intraband transition to be in the order of $10^4$ cm$^{-1}$, which is comparable to the value of the interband absorption coefficient. Additionally, we determine the size-dependence of the oscillator strength of the intraband transition. We demonstrate a negative dependence of the intraband energy with temperature, in contrast to the positive dependence of the interband transition.**

Quantum dots (QDs) are 0-dimensional solids that exhibit atomic-like electronic structures, with discrete electronic levels separated by a true zero density of states. The formidable control in size and composition of colloidal quantum dots (CQDs) has enabled experimental study of fundamental quantum physics as well the utilization of these materials for a wide variety of applications in the fields of light emission[1] and detection,[2] photovoltaics,[3] and field effect transistors.[4]

For semiconductor (SC) device engineering, one key advantage that CQDs offer compared to conventional bulk materials is band gap tunability. The band gap of CQDs can be increased from the value of their respective bulk counterpart in a precisely controlled manner, by decreasing the size of the dots. This characteristic allows QDs of *intermediate* band-gap bulk SCs, such as cadmium chalcogenides, to be tuned in the entire range of visible light;[5] QDs of *low* band- gap SCs, such as lead chalcogenides, to span over the near or short-wave infrared (NIR or SWIR);[6] and QDs of semimetals, such as mercury chalcogenides, to span over the mid- and long-wave infrared (MWIR and LWIR).[7]

In recent years, yet another degree of freedom in the band gap tunability of CQDs is being explored. By heavily populating, via doping, either the conduction band (CB) with electrons or the valence band (VB) with holes, it is possible to access optically *intraband transitions* in those bands.[8] The emergence of intraband absorption is coupled to progressive bleaching of interband absorption, as the CB or VB become more populated. Intraband transitions, taking place between electronic states within a band, are less energetic than the conventionally exploited interband transitions, and can even be smaller than band gap of the bulk counterpart, thus allowing exploitation of part of the electromagnetic spectrum forbidden for a given bulk SC material. Just to give an example of its potential, fine tuning of and access to intraband transitions in CQDs could lead to photodetectors in the MWIR and LWIR[9] ranges made of relatively high bandgap, cheap and non-toxic materials, replacing the current technology based on epitaxially-grown HgCdTe, InGaAs and GaInSb.[10] Likewise, one could think of developing cheap light sources (light-emitting diodes or lasers) covering those spectral regions. Consequently, work in this field is becoming more and more intense, and steady-state probing of intraband transitions has already been achieved in several different materials.[11] However, so far only mercury chalcogenide CQDs have exhibited stable heavy-doping that allowed fabrication of the first intraband photodetectors.[9,12] Recently, robust heavy doping has been demonstrated in PbS CQDs,[13] enabling intraband absorption and photodetection in the MWIR and LWIR ranges,[14] at energies smaller than the PbS bulk band gap. At this point, studies of the basic properties of intraband transitions in CQDs are needed to guide the development on new intraband-based devices and applications. In this work, we investigate the size and temperature dependence of intraband transitions in the CB of PbS CQDs. We also measure the intraband absorption coefficient of PbS QD films and the size dependence of the oscillator strength of the first intraband transition in PbS QDs.

**PbS CQDs and doped films used for the study**

**Figure 1** shows absorbance spectra of the PbS CQDs used in this study, which evidence good control on the QD size (diameter) in the 5–9 nm range and low size dispersion. QD size has been obtained using the empirical model for oleic-acid capped QDs reported in Ref. 6, after fitting the exciton energy in the absorbance spectra.

After synthesis, we fabricated doped films following the method reported in Ref. 13. The films are spin-coated layer-by-layer on silicon substrates, exchanging the original oleic-acid ligand with iodine. We use semi-insulating (low-doped) silicon substrates throughout the whole study because the different measurements performed require transparency in the 1–12 μm range. After formation, the film is infiltrated and capped with alumina by atomic layer deposition. The resulting films are heavily n-doped[13] and, consequently, exhibit steady-state intraband absorption in the CB.[14] **Figure 2**b shows the first two interband excitons in the doped films as a function of the dot diameter, $d$, obtained via analysis of the absorption spectra, as shown in **Figure 2**a. We found that the first exciton is red-shifted in the doped films as compared to the as-synthesized solutions (Supplementary Figure S1). We have used this exciton-shift fingerprint as well as the doping level of the samples (discussed later and shown in **Figure** 4d) to discard oxidation of the dots before or during film formation, thus guaranteeing proper sample quality for further quantitative studies. We found that the samples that showed signs of oxidation (slight blue-shift of the 1st exciton) also exhibited lower doping levels.

We have fitted the first exciton, $E_1$ (1S$_h$-1S$_e$), and the second exciton, $E_2$ (1S$_h$-1P$_e$), with Eqs. 1 and 2.

$$E_1(\text{eV}) = 0.41 + \frac{1.50}{d} + \frac{4.2}{d^2} \qquad (1)$$

with a fit correlation coefficient, $R^2 > 0.99$.

$$E_2(\text{eV}) = 0.41 + \frac{3.09}{d} + \frac{1.4}{d^2} \qquad (2)$$

with $R^2 > 0.99$.

The constant term 0.41 eV is the bulk band gap of PbS, $E_g$, at room temperature (RT). The size-dependent terms $1/d$ and $1/d^2$ represent the Coulomb interaction and quantum confinement energies, respectively. In both transitions, the Coulomb interaction term is dominant over the quantum confinement one, as it is the case in as synthesized PbS CQDs.[6] In fact, for $E_2$, data can be fitted without the $1/d^2$ term in the studied size range ($d$ = 5–8 nm) with $R^2 > 0.99$. However, the fact that the $1/d^2$ term of $E_2$ is almost negligible in this size range is not particular of doped dots. Cademartiri et al.[15] showed that the 2nd exciton in undoped PbS CQDs dispersed in solution deviates from the expected behavior for $d > 5$ nm, which causes that $E_2$ cannot be fitted to an equation of the form of Eq. 1 in the 4–7 nm size range. Therefore, although we expect Eq. 1 to be valid for $d < 5$ nm,[6,15] we do not expect the same for Eq. 2.

There is not a clear explanation for the peculiar size-dependence of $E_2$. Although it was a cause of controversy in the past, there is now consensus that it corresponds to the $1S_h$-$1P_e$ and $1P_h$-$1S_e$ transitions.[16,17] These two have very similar energies due to the almost identical electron and hole effective masses in PbS.[18] Both transitions should be forbidden by symmetry. However, it has been shown that parity selection rules are relaxed in PbS QDs, allowing optical absorption features that do not conserve parity.[16] It has been discussed that breaking of the inversion symmetry of the wavefunction is needed to fully explain the optical absorption spectra of PbS QDs.[17] Ref. 15 shows yet another higher-energy transition (7th exciton) that exhibits a size-dependence behavior similar to $E_2$, with a sudden change in trend for $d > 5$ nm, and discusses the possibility that this transition originates from a separate corner of the Brillouin zone. Here, we propose a new explanation for the sudden change in the size-dependence of $E_2$. It has been reported that the morphology of PbS QDs changes with size, evolving from an octahedral shape, for $d < 3$ nm, to a cuboctahedral shape, for $d > 4$ nm.[19] Small QDs have Pb-rich (111) facets, while in larger dots S-rich (100) facets emerge. We argue that the change in morphology in large dots, having exposed sulfur atoms, may contribute to the formation of new states that modify the energy of the 2nd (and possibly the 7th) interband transition.

**Size dependence of the intraband transition**

We have measured the absorption, *A*, spectra of the intraband transitions in our films by means of transmission and reflection measurements via Fourier transform infrared spectroscopy. Our samples consist of a thin film (~ 100 nm) of PbS QDs on top of a substrate (~ 300 μm) of semi-insulating silicon, as sketched in Supporting Figure S2. *A* is defined as:

$$A = 1 - R_{meas} - T_{meas} - A_{subs} \tag{3}$$

where $R_{meas}$ and $T_{meas}$ are the measured transmission and reflection, and $A_{subs}$ is the absorption of the substrate, which was previously characterized through reflection and transmission measurements (see Supporting Figure S3). **Figure 3**a shows that the intraband transition shifts to higher energies as the dot size decreases, owing to increased quantum confinement. Sharp features at around 0.14 eV and 0.15 eV arise from the differences (such as the amount of $SiO_2$) in the reference silicon substrate and the substrate of each sample (see Supplementary Figure S4). Below 0.13 eV, alumina contributes to absorption (Supporting Figure S5).

The size dependence of the intraband energy, $E_{IB}$, at RT is plotted in **Figure 3**b. We have fitted the results with Eq. 4.

$$E_{IB}(\text{eV}) = \frac{1.57}{d} - \frac{2.5}{d^2} \tag{4}$$

with $R^2 = 0.97$. The negative value of the $1/d^2$ term is expected, since we expect $E_{IB}$ to be equal to $E_2 - E_1$ (see Eqs. 1 and 2). For comparison, we have plotted $E_2 - E_1$ in **Figure 3**b. These results support the fact that, for all studied samples, the absorption peak analyzed correspond to intraband transitions. As it was the case for $E_1$ and $E_2$, the dominant contribution to the size dependence of $E_{IB}$ comes from Coulomb interactions. This will be useful later on for the analysis of the temperature dependence of $E_{IB}$. As discussed previously, Eq. 2 is not valid for *d* < 5 nm, therefore, Eq. 4 should not be valid either. We do not know whether Eq. 2, and hence Eq. 4, still hold for *d* > 8.5 nm. We know, however, that $E_{IB}$ should get progressively closer to zero as quantum confinement gets weaker. For dot diameters close to the exciton Bohr radius, $a_B$, of PbS (~ 20 nm), the nanocrystals are in the weak confinement regime and $E_{IB}$ should approach zero.

In the framework of the effective mass approximation, the normalized confinement energy of the different interband transitions (confinement energy of exciton $i$, $E_i - E_g$, divided by the confinement energy of the first exciton, $E_1 - E_g$) should be independent of $E_1 - E_g$.[20] **Figure 3c** plots the normalized confinement energy of the second intraband transition. We have used two set of values for $E_2$, those obtained directly from absorption measurements (**Figure 2**b), and values obtained as $E_1 + E_{IB}$ (**Figure 2**b and **Figure 3**b). As expected, both set of values yield very similar results. However, instead of a size-independent value, we observe a linear decrease of the normalized confinement energy of the second exciton for $E_1 - E_g$ in the 0.25–0.45 eV range (8 nm > $d$ > 5 nm). This behavior matches the one reported for undoped PbS CQDs dispersed in solution,[15] where $(E_2 - E_g)/(E_1 - E_g) \approx 1.39$ for $E_1 - E_g > 0.45$ eV ($d$ < 5 nm) and increases for smaller confinement energies (larger dots). We ascribe this behavior to the fact that both $E_1$ an $E_2$ have a predominant contribution of the Coulomb interaction rather than the $1/d^2$ confinement energy term, as discussed in the previous section.

**Figure 3**c also shows measurements of $(E_2 - E_g)/(E_1 - E_g)$ at 98 K. To calculate $E_g$ at low temperatures, we have used 320 μeV/K as the temperature dependence of the bulk PbS band gap.[21] The low temperature values of normalized confinement energies match closely the values at RT, indicating the $E_1$ and $E_2$ at low temperatures follow similar size dependences to those shown in **Figure 2**b. The impact of temperature on the intraband transition will be studied in the next section.

**Absorption coefficient of doped PbS QD films**

Characterizing the intrinsic absorption intensity of a semiconductor is crucial both for its use in practical applications and for theoretical studies. In bulk semiconductors the absorption coefficient, $\alpha$, measured in units of inverse length (cm$^{-1}$), is used. For semiconductor nanoparticles dispersed in a solution, the molar extinction coefficient, $\varepsilon$, measured is units of inverse molarity and length (mol$^{-1}$·cm$^{-1}$), is usually employed instead, since absorption obtained experimentally depends not only on the path length, but also on the particle concentration, as defined by Beer-Lambert law. A theoretical intrinsic absorption coefficient for dispersed nanoparticles, $\mu_i$, can be obtained from $\varepsilon$, if the particle volume is known.[22]

In our case, we have a thin-film of densely packed nanoparticles, therefore, the determination of $\alpha$ is pertinent for evaluating the potential of heavily doped PbS QDs for intraband optoelectronics. We have performed absorption measurements such as those shown in **Figure** 3a to calculate the absorption coefficient of our material. For the simple case of a thin slab of homogenous material, $\alpha$ can be calculated exactly by taking into consideration the multiple internal reflections that occur at the front and back air/sample interfaces, which increase the effective optical path.[23] When the thin-film is deposited on a substrate, reflections taking place at every interfaces (refractive-index change), as well as absorption in the substrate, must be taken into account. For this case, $\alpha$ can also be analytically calculated assuming $\alpha t \ll 1$,[24] where $t$ is the thickness of the thin film. However, in all cases the refractive index of the sample or, alternatively, the fraction of reflected light in a single air/sample interface, has to be known. Because of the wavelength range in our study (5–9 μm), it was not possible for us to measure either of these values. Hence, we have calculated $\alpha$ using a simple model described by Eq. 5:

$$A = 1 - e^{-\alpha t} \quad (5)$$

Eq. 5 does not take into account the multiple reflections taking place inside the sample and, therefore, underestimates the effective optical path. For this reason, our results must be understood as an *upper bound* to the absorption coefficient of the intraband transition in doped PbS QDs. Note that $A$ is obtained after correcting for the absorption of the substrate (Eq. 3 and Supplementary Figure S3). **Figure 4**a shows some of the measured absorption coefficient spectra of the intraband region. The peak values of $\alpha$ for the $1S_e$-$1P_e$ transition are plotted in **Figure 4**b as a function of $d$. The values are in the order of $10^4$ cm$^{-1}$, similarly to the values obtained for the $1S_h$-$1S_e$ interband transition in undoped PbS QDs.[22] $\alpha$ peak values and other relevant parameters of the results presented in **Figure 4** are compiled in Supplementary Table S1. To make a better comparison between absorption intensity in different films, **Figure 4**c shows the values of $\alpha$ integrated over the intraband absorption range, $\alpha_{IB}$. Integration over the whole spectrum of the intraband transition corrects for the impact on $\alpha$ peak values of differences in full width half maxima (FWHM, see Supplementary Table S1) of the intraband absorption spectra. We ascribe the differences on FWHM to different size distribution in the synthesized nanoparticles.

In our QD films, $\alpha$ and $\alpha_{IB}$ are influenced by factors extrinsic to the QDs, such as the packing density of the films (or the volume fraction occupied by the QDs) and the presence of a surrounding medium (alumina), imposed by the doping procedure, that alters the effective dielectric function of the film.[22] For fundamental studies that require comparison with theory, the oscillator strength is a more adequate parameter. The oscillator strength is an intrinsic property of materials, related to the probability of an electronic quantum transition upon the presence of a resonant electromagnetic field. The integrated absorption coefficient of the QD film is directly related to the oscillator strength *per volume*, $f_V$.[25] The oscillator strength *per particle*, $f_{QD}$, instead, determines the radiative lifetime. Since we ignore the dielectric constant of our material in the energy range of our study, as well as the exact absorption coefficient, we cannot make a quantitative study of the oscillator strength of our particles. However, under the assumption that the film packing density and dielectric constant are similar in all samples, we are able to evaluate in qualitative terms the dependence of $f_{QD}$ with $d$. Before we do it, one last aspect has to be considered: the doping level of our samples. As discussed above, more sulfur sites are exposed at the surface as PbS QDs get larger. It is precisely the presence of exposed sulfur sites that allows heavy doping in the employed doping method.[13] Hence, the doping level of our QDs vary with size. PbS QDs can accommodate up to 8 electrons in the $1S_e$ state.[18] We define the $1S_e$ occupancy factor $\emptyset = n_{QD}/8$, where $n_{QD}$ is the doping level of the QD films (measured in electrons/QD). **Figure 4**d shows the measured $\emptyset$ as a function of $d$. While the oscillator strength defines the probability of a quantum absorption process to take place, the occupancy of the initial and final states has to be additionally considered to calculate the absorption coefficient.[26] Thus, in undoped PbS QDs (empty $1S_e$) $\alpha_{IB} = 0$, because the initial state of the intraband transition is empty, while $f_{QD} \neq 0$, because it is independent of the occupancy of the initial and final states. Therefore, for a proper comparison of the oscillator strength in different samples, $\alpha_{IB}$ has to be corrected by $\emptyset$. $\alpha_{IB}/\emptyset$ determines the maximum attainable value of $\alpha_{IB}$, achieved at full population of $1S_e$. This correction allows us to relate directly the measured $\alpha_{IB}$ to $f_V$, as shown in **Figure 4**e. Note that, implicitly, we are assuming that –resulting from Fermi-Dirac statistics– the $1P_e$ state is virtually empty at RT for all samples, given $E_{IB} > 0.15$ eV. $f_{QD}$ is calculated as $f_{QD} = f_V \times V$, where $V$ is the volume of a QD of diameter $d$. **Figure 4**f plots the size dependence of $f_{QD}$, assuming the QDs to be spherical. We have fitted $f_V$ to a power law equation, resulting in an exponent value of 2.7 and $R^2 = 0.93$. This yields to a power law fit with exponent 0.3 for $f_{QD}$, which is shown in **Figure 4**f as a guide to the eye. Our results reveal a weak dependence with size of the oscillator strength per particle of the $1S_e$-$1P_e$ transition. This will be further discussed in the last section of the article.

**Temperature dependence of the intraband transition**

Analysis of the temperature dependence of the interband and intraband transitions is also fundamental to enable the development of theoretical models that can predict the optical properties of CQD materials. For example, applications such as lighting or photodetection are sensitive to chromatic drift. The temperature dependence of the interband transition in cadmium and lead chalcogenide CQDs has been extensively analyzed.[27–33] Note that some of these works use the notation $E_g$ for the 1$^{st}$ interband transition, while we reserve it for the bulk band gap in this article.

In the case of PbS, Olkhovets et al.[27] showed that $E_1$ is nearly independent of temperature in small dots ($d$ < 4 nm), as expected for atomic-like energy levels. As the dots get larger, $E_1$ has an increasing positive dependence with temperature, approaching the temperature dependence of bulk PbS as quantum confinement is reduced. Different theoretical models have been used to explain the experimental results, taking into account the impact of lattice expansion,[27] electron-phonon (el-ph) interactions,[27,28,30] and effective mass.[31] From these works, it is concluded that el-ph interactions play a predominant role in the temperature dependence of $E_1$.

We have measured the intraband absorption of our heavily doped samples in the temperature range 98–298 K (see Methods). **Figure 5**a-b show measurements for two samples with different dot size. The first noticeable feature is the red-shift of $E_{IB}$ as temperature increases, conversely to the reported blue-shift of $E_1$ in undoped PbS CQDs. The positive dependence of $E_1$ with temperature is preserved in doped QDs, as shown in **Figure 5**c. Another interesting aspect is the apparent increase in intraband absorption intensity upon sample cooling. We think that this is an intrinsic property of the material, not due to a variation of $n_{QD}$ with temperature, since the doping level of the dots has been shown to be constant in the 80–300 K range.[14] The increase of intraband absorption intensity at low temperatures is in line with previous reports of increased absorption cross section of the interband transition at low temperatures in PbS CQDs.[29] Finally, by comparing **Figure 5**a and b, it can be appreciated that the energy shift of the intraband transition is stronger for smaller dots, as opposed to what has been found for the interband case.

In order to do a quantitative analysis of the temperature dependence of the intraband and interband transitions, we have identified the energies $E_1$ and $E_{IB}$ for all temperatures, as described in previous sections. In addition, we have measured samples with light n-type doping with $n_{QD} \ll 1$ (see Methods) in which there are no signs of interband bleaching or intraband absorption. An example of temperature dependence for such samples is shown in Supporting Figure S6. For $E_1$ and $E_{IB}$, we have calculated the respective coefficients of energy variation with temperature, $\partial E_1/\partial T$ and $\partial E_{IB}/\partial T$, as the slopes of a linear fits to the measured data (Supporting Figure S7). Although the temperature dependence of $E_1$ is not linear in the whole 0–300 K range,[29,30] the use of a linear dependence is generally accepted as a valid approximation in the 100–300 K range.[27,29] Our results show that a linear dependence is also a good approximation for the intraband transition in that temperature range.

The measured temperature dependence coefficients are plotted in **Figure 5**d as a function of the dot size. As anticipated, the temperature dependence of the intraband transition is negative and it is stronger for smaller dots. These results can be qualitatively explained by considering only el-ph coupling effects, which, as mentioned before, are dominant in the temperature dependence of $E_1$ in undoped PbS QDs. The contribution of el-ph coupling to the temperature dependence of interband transitions in PbS QDs is smaller for higher energies.[27] Thus, the temperature dependence of $E_2$ is expected to be weaker than that of $E_1$. Since $E_{IB} = E_2 - E_1$, it follows that $\partial E_{IB}/\partial T < 0$. In addition, since the energy difference between $E_2$ and $E_1$ is larger for smaller dots, $\partial E_{IB}/\partial T$ should increase (in magnitude) as the dot size is reduced.

In our lightly doped QD films, the values of $\partial E_1/\partial T$ are similar to those previously reported for undoped QDs.[27] Upon heavy doping, $\partial E_1/\partial T$ is reduced. The difference in $\partial E_1/\partial T$ between the heavily doped and lightly doped samples increases with dot size, and, hence, with ∅ (see **Figure 4**d). For samples with $d = 5$ nm (∅~0.1) both values are very close. We deduce that upon occupation of 1S$_e$ the el-ph interaction is modified, resulting in a reduced contribution to the temperature dependence of $E_1$.

Next, we analyze the size dependence of $\partial E_{IB}/\partial T$ on the basis of empirical results discussed above. We know, through Eqs. 1 and 2, that at RT both $E_1$ and $E_2$ are dominated by the Coulomb term (1/$d$). In addition, **Figure 3**c indicates that this behavior is maintained at low temperatures. The prefactor of the 1/$d$ term is different for each transition because it depends on the wavefunctions of the electron and hole.[34] Thus, we will approximate $E_1$ and $E_2$ to

$$E_n(\text{eV}) = E_g + \frac{A_n}{d} \tag{6}$$

where the $n = 1, 2$ and $A_n$ is the prefactor of the Coulomb term of the $n$-th interband transition. Neglecting lattice expansion effects, we get:

$$\frac{\partial E_{IB}}{\partial T} = \frac{\partial (E_2 - E_1)}{\partial T} = \left(\frac{\partial A_2}{\partial T} - \frac{\partial A_1}{\partial T}\right)\frac{1}{d} \tag{7}$$

For simplicity, we will assume that $\partial A_2/\partial T - \partial A_1/\partial T = C$; i. e., that the temperature dependences of $A_2$ and $A_1$ are related by a constant factor $C$. This would be the case if, for example, we assume that the permittivity of the QDs, $\varepsilon$, is the only parameter in $A_n$ that is dependent on temperature and we approximate $\partial \varepsilon/\partial T$ to a constant.[35] Then, Eq. 7 turns into Eq. 8:

$$\frac{\partial E_{IB}}{\partial T} = \frac{C}{d} \tag{8}$$

**Figure 5**d shows that the measured data of $\partial E_{IB}/\partial T$ fits reasonably well to Eq. 8 ($R^2 = 0.96$). For large values of $d$, approaching $a_B$, $\partial E_{IB}/\partial T$ approaches zero, as it should, since $E_{IB} \to 0$ because of the relaxation of the quantum confinement.

**Discussion and conclusions**

Robust heavy n-doping of the CB leads to steady-state intraband absorption in PbS CQDs in densely packed films. The fact that our experiments have been performed on thin films instead of solutions introduces limitations as well as advantages. On one hand, it complicates the exact calculation of fundamental parameters such as the absorption coefficient or oscillator strength, due to the necessity of knowing the QD volumetric concentration in the film for each QD size, as well as the dielectric constant of the surrounding material. On the other hand, because we have studied QD solids, our results provide valuable data that can be directly used in device design and engineering.

We have demonstrated that the absorption coefficient of the intraband transition is in the order of the absorption coefficient of the interband transition ($10^4$ cm$^{-1}$). This result implies that thin-film PbS intraband optoelectronics can be envisaged. Additionally, dots with semi-occupied $1S_e$ (only partially bleached interband) exhibit strong absorption in both interband and intraband transitions. These dual-band strong absorbers may be interesting for multi-gap optoelectronic devices, such as intermediate band solar cells,[36] which currently experience a major setback related to weak absorption.[37]

Our measurements reveal absorption in the first intraband transition ($1S_e$-$1P_e$). However, absorption at more energetic intraband transitions (from $1S_e$ to higher levels such as $1D_e$, $2S_e$, $1F_e$[16]) cannot be appreciated (Supporting Figure S8). We conclude that these transitions are either forbidden or much weaker than the first intraband transition in the studied size range.

The absorption strength of interband transitions has been studied for different semiconductor nanocrystals by measuring the absorption coefficient, the molar extinction coefficient or the absorption cross section.[6,15,33,38–43] Early calculations on the strength of interband transitions suggested that $f_{QD}$ for $E_1$ should be independent of the dot size, so long the dots are in the strong confinement regime.[34] However, it has been pointed out that further theoretical studies, going beyond the effective mass approximation, are needed to explain the experimental results obtained for different materials.[22] Our work indicates that the oscillator strength per particle of the intraband transition also has a weak dependence on $d$. Some theoretical works have calculated the oscillator strength of intraband transitions for QDs of different materials.[26,44,45] However, because of the particularities of each study, none of these works can be directly related to our case. We hope that the experimental results presented herein will motivate theoretical studies of the oscillator strength of intraband transitions in lead chalcogenide materials.

We have provided simple analysis, based on empirical results, of the temperature dependence of $E_{IB}$ for PbS QDs. A more fundamental and general theoretical framework for the temperature dependence of intraband transitions is needed. It can be addressed by analyzing the temperature dependence of $E_2$ and $E_1$ considering the occupancy of $1S_e$, which has a noticeable impact in them.

Finally, we would like to remark that this study has been conducted for the first electron intraband transition, due to the n-type character of the dots. Because of the almost identical electron and hole effective masses of PbS, corresponding results for the first hole intraband transitions (in a p-type material) are expected to be similar.

METHODS

**PbS colloidal quantum dot synthesis.** PbS QDs synthesis was adapted from a previously reported multi-injection procedure.[46] Briefly, a mixture of lead oxide (PbO), 1-octadecene (ODE) and oleic acid was degassed overnight at 90 °C under vacuum. After degassing, the solution was placed under Ar atmosphere and a specific reaction temperature was set. A solution of hexamethyldisilathiane ((TMS)$_2$S) in ODE was quickly injected. After 6 minutes, a second solution of (TMS)$_2$S in ODE was dropwise injected in a rate of 0.75 mL/min. After this second injection, the heating was stopped and the solution was let to naturally cool down to room temperature. QDs were purified three times by precipitation with acetone and ethanol and redispersed in anhydrous toluene. Finally, the concentration was adjusted to 30 mg/mL and the solution was bubbled with N$_2$. Depending on the reaction parameters, PbS QDs of different sizes were synthesized.

**n-doped PbS quantum-dot film fabrication.** PbS CQD films were deposited using a layer-by-layer spin-coating process under ambient conditions at 2500 r.p.m. For each layer, the CQD solution was deposited on semi-insulating (10-20 Ohm·cm) double-polished Si. Solid-state ligand exchange was performed by flooding the surface with 1-ethyl-3-methylimidazolium iodide in methanol (EMII, 7 mg/ml) 30 s before spin-coating dry at 2500 r.p.m. Two washes with methanol were used to remove unbound ligands. After film formation, samples are lightly n-doped, with the 1S$_e$ state practically unpopulated.[13] Heavy n-doping is achieved by infiltrating and capping the films with alumina (Al$_2$O$_3$) via atomic layer deposition (ALD), following an already reported method.[13]

**Measurement of the QD doping level.** A baseline correction was applied to the absorption measurements in order to allow proper comparison between films before and after the ALD process. Since the 1S$_e$ states of PbS are eight fold degenerated (including spin), the number of electrons per dot in the CB, $n_{QD}$, can be calculated in a straightforward manner from the bleach of the first exciton transition. If we define $I_1$ and $I_2$ as the integrated absorption strength of the excitonic transition of the lightly doped and heavily doped samples, respectively, then $n_{QD} = 8(1 - I_2/I_1)$. Note that we are assuming that in the lightly doped samples the doping level of the samples $n_{QD} \ll 1$ so that we can consider full valence band and empty conduction band. For related figures see Ref. 13.

**Absolute absorption measurements.** Absorption was obtained by means of reflection and transmission measurements, as described in the main text. Room-temperature transmission and reflection measurements were made using a Cary 600 FTIR with microscope. The bench of the FTIR was purged with nitrogen gas to minimize the impact of atmospheric absorption in the measurements.

**Temperature-dependent absorption measurements.** Absorption was obtained as: 1 – transmission. Temperature variable transmission measurement were made under vacuum, using a Cary 610 FTIR with microscope coupled to a liquid-nitrogen cooled, temperature-controllable Linkam HFS350EV-PB4 stage equipped with ZnSe windows. For all samples, the temperature change was done at a controlled rate (3 K/min). After reaching the temperature set point, temperature was stabilized during 5 minutes prior to performing the measurements.

A bare-silicon substrate was used as background sample. The same background, measured at RT, was used to determine the transmission of the samples at all temperatures. Low-temperature measurements exhibit a double peak feature centered at 0.136 eV. This feature is due to the temperature-dependent energy shift of the $SiO_2$ absorption in the samples, which is not adequately corrected by the RT background.

**Determination of sample thickness.** The thicknesses of the samples used in absorption measurement were measured using a KLA-Tencor Alpha-Step IQ Surface Profiler.

ASSOCIATED CONTENT

**Supporting Information**
Supporting Figures S1–S8 and Table S1.

AUTHOR INFORMATION

**Corresponding Authors**
*Address correspondence to: i.ramiro@ies.upm.es; Gerasimos.Konstantatos@icfo.eu

**Present Address**
I. Ramiro's present address is: Instituto de Energía Solar, Universidad Politécnica de Madrid, Avda. Complutense 30, 28040, Madrid, Spain.

**Author Contributions**
I. R. designed the experiments and analyzed the results. I. R., B. K., and M. P. performed the absorption measurements. M. D. synthesized the QDs. I. R. and O. Ö. fabricated the samples. I. R. and G. K. wrote the manuscript with input from all authors. G. K. supervised the work.

ACKNOWLEDGEMENTS

I. R. acknowledges Dr. Iacopo Torre for helpful discussions. The authors acknowledge financial support from the European Research Council (ERC) under the European Union's Horizon 2020 research and innovation programme (grant agreement no. 725165), the Spanish Ministry of Economy and Competitiveness (MINECO), and the "Fondo Europeo de Desarrollo Regional" (FEDER) through grant TEC2017-88655-R. The authors also acknowledge financial support from Fundacio Privada Cellex, the program CERCA and from the Spanish Ministry of Economy and Competitiveness, through the "Severo Ochoa" Programme for Centres of Excellence in R&D (SEV-2015-0522).

REFERENCES

(1) Shirasaki, Y.; Supran, G. J.; Bawendi, M. G.; Bulović, V. Emergence of Colloidal Quantum-Dot Light-Emitting Technologies. *Nat. Photonics* **2013**, *7* (1), 13–23. https://doi.org/10.1038/nphoton.2012.328.

(2) Konstantatos, G.; Sargent, E. H. Colloidal Quantum Dot Photodetectors. *Infrared Phys. Technol.* **2011**, *54* (3), 278–282. https://doi.org/10.1016/J.INFRARED.2010.12.029.

(3) Carey, G. H.; Abdelhady, A. L.; Ning, Z.; Thon, S. M.; Bakr, O. M.; Sargent, E. H. Colloidal Quantum Dot Solar Cells. *Chem. Rev.* **2015**, *115* (23), 12732–12763. https://doi.org/10.1021/acs.chemrev.5b00063.

(4) Hetsch, F.; Zhao, N.; Kershaw, S. V.; Rogach, A. L. Quantum Dot Field Effect Transistors. *Mater. Today* **2013**, *16* (9), 312–325.

https://doi.org/10.1016/j.mattod.2013.08.011.

(5) Murray, C. B.; Norris, D. J.; Bawendi, M. G. Synthesis and Characterization of Nearly Monodisperse CdE (E = S, Se, Te) Semiconductor Nanocrystallites. *J. Am. Chem. Soc.* **1993**, *115* (19), 8706–8715. https://doi.org/10.1021/ja00072a025.

(6) Moreels, I.; Lambert, K.; Smeets, D.; De Muynck, D.; Nollet, T.; Martins, J. C.; Vanhaecke, F.; Vantomme, A.; Delerue, C.; Allan, G.; et al. Size-Dependent Optical Properties of Colloidal PbS Quantum Dots. *ACS Nano* **2009**, *3* (10), 3023–3030. https://doi.org/10.1021/nn900863a.

(7) Keuleyan, S. E.; Guyot-Sionnest, P.; Delerue, C.; Allan, G. Mercury Telluride Colloidal Quantum Dots: Electronic Structure, Size-Dependent Spectra, and Photocurrent Detection up to 12 Mm. *ACS Nano* **2014**, *8* (8), 8676–8682. https://doi.org/10.1021/nn503805h.

(8) Shim, M.; Guyot-Sionnest, P. N-Type Colloidal Semiconductor Nanocrystals. *Nature* **2000**, *407* (6807), 981–983. https://doi.org/10.1038/35039577.

(9) Jagtap, A.; Livache, C.; Martinez, B.; Qu, J.; Chu, A.; Gréboval, C.; Goubet, N.; Lhuillier, E. Emergence of Intraband Transitions in Colloidal Nanocrystals [Invited]. *Opt. Mater. Express* **2018**, *8* (5), 1174. https://doi.org/10.1364/OME.8.001174.

(10) Rogalski, A. Recent Progress in Infrared Detector Technologies. *Infrared Phys. Technol.* **2011**, *54* (3), 136–154. https://doi.org/10.1016/j.infrared.2010.12.003.

(11) Kim, J.; Choi, D.; Jeong, K. S. Self-Doped Colloidal Semiconductor Nanocrystals with Intraband Transitions in Steady State. *Chem. Commun.* **2018**, *54* (61), 8435–8445. https://doi.org/10.1039/c8cc02488j.

(12) Deng, Z.; Jeong, K. S.; Guyot-Sionnest, P. Colloidal Quantum Dots Intraband Photodetectors. *ACS Nano* **2014**, *8* (11), 11707–11714. https://doi.org/10.1021/nn505092a.

(13) Christodoulou, S.; Ramiro, I.; Othonos, A.; Figueroba, A.; Dalmases, M.; Özdemir, O.; Pradhan, S.; Itskos, G.; Konstantatos, G. Single-Exciton Gain and Stimulated Emission Across the Infrared Optical Telecom Band from Robust Heavily-Doped PbS Colloidal Quantum Dots. *arXiv:1908.03796 [physics.app-ph]* **2019**.

(14) Ramiro, I.; Özdemir, O.; Christodoulou, S.; Gupta, S.; Dalmases, M.; Torre, I.; Konstantatos, G. Mid- and Long-Wave Infrared Optoelectronics via Intraband Transitions in PbS Colloidal Quantum Dots. *Nano Lett.* **2020**.

(15) Cademartiri, L.; Montanari, E.; Calestani, G.; Migliori, A.; Guagliardi, A.; Ozin, G. A. Size-Dependent Extinction Coefficients of PbS Quantum Dots. *J. Am. Chem. Soc.* **2006**, *128* (31), 10337–10346. https://doi.org/10.1021/ja063166u.

(16) Diaconescu, B.; Padilha, L. A.; Nagpal, P.; Swartzentruber, B. S.; Klimov, V. I. Measurement of Electronic States of PbS Nanocrystal Quantum Dots Using Scanning Tunneling Spectroscopy: The Role of Parity Selection Rules in Optical Absorption. *Phys. Rev. Lett.* **2013**, *110* (12), 127406. https://doi.org/10.1103/PhysRevLett.110.127406.

(17) Nootz, G.; Padilha, L. A.; Olszak, P. D.; Webster, S.; Hagan, D. J.; Van Stryland, E.

W.; Levina, L.; Sukhovatkin, V.; Brzozowski, L.; Sargent, E. H. Role of Symmetry Breaking on the Optical Transitions in Lead-Salt Quantum Dots. *Nano Lett.* **2010**, *10* (9), 3577–3582. https://doi.org/10.1021/nl1018673.

(18) Kang, I.; Wise, F. W. Electronic Structure and Optical Properties of PbS and PbSe Quantum Dots. *J. Opt. Soc. Am. B* **1997**, *14* (7), 1632. https://doi.org/10.1364/JOSAB.14.001632.

(19) Beygi, H.; Sajjadi, S. A.; Babakhani, A.; Young, J. F.; van Veggel, F. C. J. M. Surface Chemistry of As-Synthesized and Air-Oxidized PbS Quantum Dots. *Appl. Surf. Sci.* **2018**, *457* (May), 1–10. https://doi.org/10.1016/j.apsusc.2018.06.152.

(20) Wehrenberg, B. L.; Wang, C.; Guyot-Sionnest, P. Interband and Intraband Optical Studies of PbSe Colloidal Quantum Dots. *J. Phys. Chem. B* **2002**, *106* (41), 10634–10640. https://doi.org/10.1021/jp021187e.

(21) Gibbs, Z. M.; Kim, H.; Wang, H.; White, R. L.; Drymiotis, F.; Kaviany, M.; Jeffrey Snyder, G. Temperature Dependent Band Gap in PbX (X = S, Se, Te). *Appl. Phys. Lett.* **2013**, *103* (26), 1–5. https://doi.org/10.1063/1.4858195.

(22) Hens, Z.; Moreels, I. Light Absorption by Colloidal Semiconductor Quantum Dots. *J. Mater. Chem.* **2012**, *22* (21), 10406–10415. https://doi.org/10.1039/c2jm30760j.

(23) Pankove, J. I. *Optical Processes in Semicondcutors*; Dover Publications, Inc, 1975.

(24) Bubenzer, A.; Koidl, P. Exact Expressions for Calculating Thin-Film Absorption Coefficients from Laser Calorimetric Data. *Appl. Opt.* **1984**, *23* (17), 2886. https://doi.org/10.1364/ao.23.002886.

(25) Wang, Y.; Herron, N. Nanometer-Sized Semiconductor Clusters: Materials Synthesis, Quantum Size Effects, and Photophysical Properties. *J. Phys. Chem.* **1991**, *95* (2), 525–532. https://doi.org/10.1021/j100155a009.

(26) De Sousa, J. S.; Leburton, J. P.; Freire, V. N.; Da Silva, E. F. Intraband Absorption and Stark Effect in Silicon Nanocrystals. *Phys. Rev. B - Condens. Matter Mater. Phys.* **2005**, *72* (15), 1–8. https://doi.org/10.1103/PhysRevB.72.155438.

(27) Olkhovets, A.; Hsu, R. C.; Lipovskii, A.; Wise, F. W. Size-Dependent Temperature Variation of the Energy Gap in Lead-Salt Quantum Dots. *Phys. Rev. Lett.* **1998**, *81* (16), 3539–3542. https://doi.org/10.1103/PhysRevLett.81.3539.

(28) Dey, P.; Paul, J.; Byisma, J.; Karaiskaj, D.; Luther, J. M.; Beard, M. C.; Romero, A. H. Origin of the Temperature Dependence of the Band Gap of PbS and PbSe Quantum Dots. *Solid State Commun.* **2013**, *165*, 49–54. https://doi.org/10.1016/j.ssc.2013.04.022.

(29) Nordin, M. N.; Li, J.; Clowes, S. K.; Curry, R. J. Temperature Dependent Optical Properties of PbS Nanocrystals. *Nanotechnology* **2012**, *23* (27). https://doi.org/10.1088/0957-4484/23/27/275701.

(30) Ullrich, B.; Wang, J. S.; Brown, G. J. Analysis of Thermal Band Gap Variations of PbS Quantum Dots by Fourier Transform Transmission and Emission Spectroscopy. *Appl. Phys. Lett.* **2011**, *99* (8), 22–25. https://doi.org/10.1063/1.3623486.

(31) Liptay, T. J.; Ram, R. J. Temperature Dependence of the Exciton Transition in Semiconductor Quantum Dots. *Appl. Phys. Lett.* **2006**, *89* (22), 223132. https://doi.org/10.1063/1.2400107.

(32) Dai, Q.; Zhang, Y.; Wang, Y.; Hu, M. Z.; Zou, B.; Wang, Y.; Yu, W. W. Size-Dependent Temperature Effects on PbSe Nanocrystals. *Langmuir* **2010**, *26* (13), 11435–11440. https://doi.org/10.1021/la101545w.

(33) Vossmeyer, T.; Katsikas, L.; Giersig, M.; Popovic, I. G.; Diesner, K.; Chemseddine, A.; Eychmüller, A.; Weller, H. CdS Nanoclusters: Synthesis, Characterization, Size Dependent Oscillator Strength, Temperature Shift of the Excitonic Transition Energy, and Reversible Absorbance Shift. *J. Phys. Chem.* **1994**, *98* (31), 7665–7673. https://doi.org/10.1021/j100082a044.

(34) Brus, L. E. Electron-Electron and Electron-Hole Interactions in Small Semiconductor Crystallites: The Size Dependence of the Lowest Excited Electronic State. *J. Chem. Phys.* **1984**, *80* (9), 4403–4409. https://doi.org/10.1063/1.447218.

(35) Sagadevan, S.; Pal, K.; Hoque, E.; Chowdhury, Z. Z. A Chemical Synthesized Al-Doped PbS Nanoparticles Hybrid Composite for Optical and Electrical Response. *J. Mater. Sci. Mater. Electron.* **2017**, *28* (15), 10902–10908. https://doi.org/10.1007/s10854-017-6869-7.

(36) Luque, A.; Martí, A.; Stanley, C. Understanding Intermediate-Band Solar Cells. *Nat. Photonics* **2012**, *6* (3), 146–152. https://doi.org/10.1038/nphoton.2012.1.

(37) Ramiro, I.; Marti, A.; Antolin, E.; Luque, A. Review of Experimental Results Related to the Operation of Intermediate Band Solar Cells. *IEEE Journal of Photovoltaics*. 2014, pp 736–748. https://doi.org/10.1109/JPHOTOV.2014.2299402.

(38) Moreels, I.; Lambert, K.; De Muynck, D.; Vanhaecke, F.; Poelman, D.; Martins, J. C.; Allan, G.; Hens, Z. Composition and Size-Dependent Extinction Coefficient of Colloidal PbSe Quantum Dots. *Chem. Mater.* **2007**, *19* (25), 6101–6106. https://doi.org/10.1021/cm071410q.

(39) Leatherdale, C. A.; Woo, W. K.; Mikulec, F. V.; Bawendi, M. G. On the Absorption Cross Section of CdSe Nanocrystal Quantum Dots. *J. Phys. Chem. B* **2002**, *106* (31), 7619–7622. https://doi.org/10.1021/jp025698c.

(40) Talapin, D. V.; Gaponik, N.; Borchert, H.; Rogach, A. L.; Haase, M.; Weller, H. Etching of Colloidal InP Nanocrystals with Fluorides: Photochemical Nature of the Process Resulting in High Photoluminescence Efficiency. *J. Phys. Chem. B* **2002**, *106* (49), 12659–12663. https://doi.org/10.1021/jp026380n.

(41) Yu, P.; Beard, M. C.; Ellingson, R. J.; Fernere, S.; Curtis, C.; Drexler, J.; Luiszer, F.; Nozik, A. J. Absorption Cross-Section and Related Optical Properties of Colloidal InAs Quantum Dots. *J. Phys. Chem. B* **2005**, *109* (15), 7084–7087. https://doi.org/10.1021/jp046127i.

(42) Karel Čapek, R.; Moreels, I.; Lambert, K.; De Muynck, D.; Zhao, Q.; Van Tomme, A.; Vanhaecke, F.; Hens, Z. Optical Properties of Zincblende Cadmium Selenide Quantum Dots. *J. Phys. Chem. C* **2010**, *114* (14), 6371–6376. https://doi.org/10.1021/jp1001989.

(43) Kamal, J. S.; Omari, A.; Van Hoecke, K.; Zhao, Q.; Vantomme, A.; Vanhaecke, F.; Capek, R. K.; Hens, Z. Size-Dependent Optical Properties of Zinc Blende Cadmium Telluride Quantum Dots. *J. Phys. Chem. C* **2012**, *116* (8), 5049–5054. https://doi.org/10.1021/jp212281m.

(44) Şahin, M. Photoionization Cross Section and Intersublevel Transitions in a One- and Two-Electron Spherical Quantum Dot with a Hydrogenic Impurity. *Phys. Rev. B - Condens. Matter Mater. Phys.* **2008**, *77* (4), 1–12. https://doi.org/10.1103/PhysRevB.77.045317.

(45) Yilmaz, S.; Şafak, H. Oscillator Strengths for the Intersubband Transitions in a CdS-SiO2 Quantum Dot with Hydrogenic Impurity. *Phys. E Low-Dimensional Syst. Nanostructures* **2007**, *36* (1), 40–44. https://doi.org/10.1016/j.physe.2006.07.040.

(46) Lee, J. W.; Kim, D. Y.; Baek, S.; Yu, H.; So, F. Inorganic UV-Visible-SWIR Broadband Photodetector Based on Monodisperse PbS Nanocrystals. *Small* **2016**, *12* (10), 1328–1333. https://doi.org/10.1002/smll.201503244.

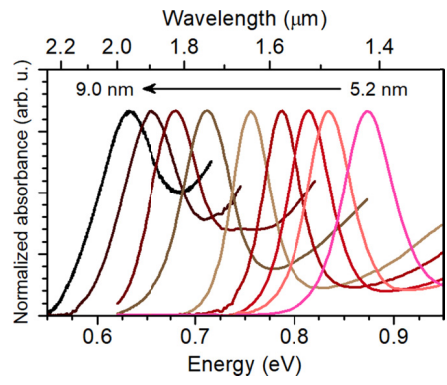

**Figure 1**. Absorbance spectra, normalized to the maximum of the exciton peaks, of oleic-acid capped PbS CQDs dispersed in toluene. The QD diameter of the different solutions ranges between 5.2 and 9.0 nm.

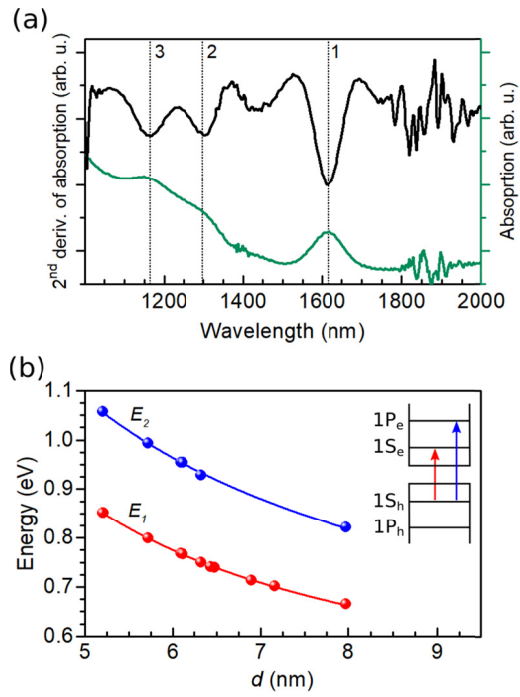

**Figure 2**. Interband absorption in heavily n-doped PbS QDs. (a) Second-derivative analysis of the absorption spectrum. Vertical lines indicate the first three excitons. (b) Energy of the first two excitons, corresponding to lines 1 and 2 in (a), as a function of $d$. The lines are fits, as described in the main text.

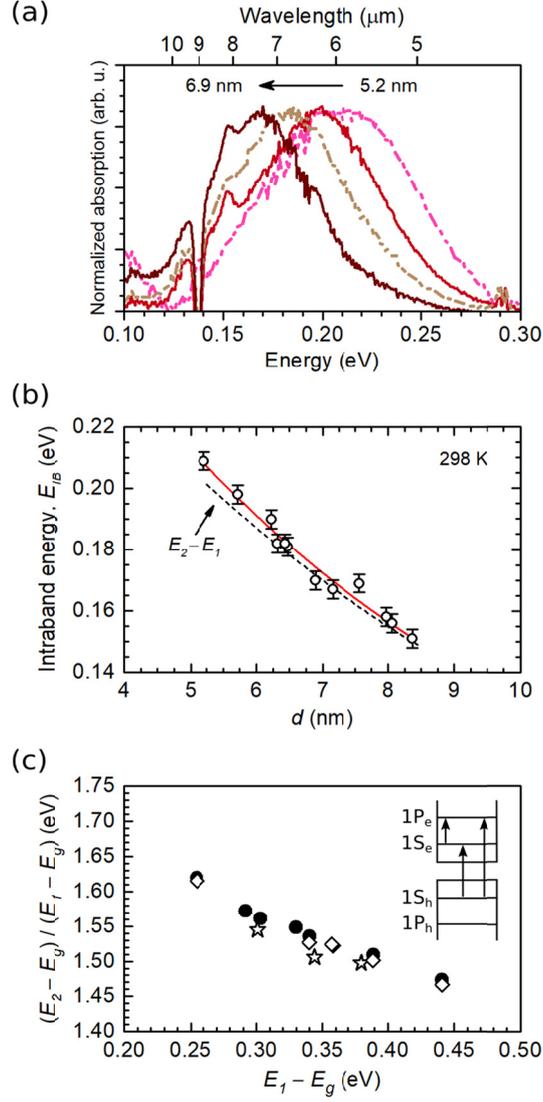

**Figure 3**. (a) Intraband absorption spectra of QDs of different sizes normalized to the peak maxima. (b) Energy of the intraband transition, at RT, as a function of the dot diameter. Data are obtained as the central energies of Gaussian fits of absorption spectra such as those presented in (a). The error bars are estimates of the uncertainty in the fits (± 3 meV). The red solid line corresponds to Eq. 4. The black dashed line represents the energy difference between the 2$^{nd}$ and 1$^{st}$ excitons. (c) Second confinement energy ($E_2 - E_g$) divided by the first confinement energy ($E_1 - E_g$) as a function of the first confinement energy. $E_2$ (1S$_h$-1P$_e$) is obtained in two ways: directly from absorption spectra at RT (◇), and as $E_{IB}$ (1S$_e$-1P$_e$) + $E_1$ (1S$_h$-1S$_e$) at RT (●) and 98 K (☆).

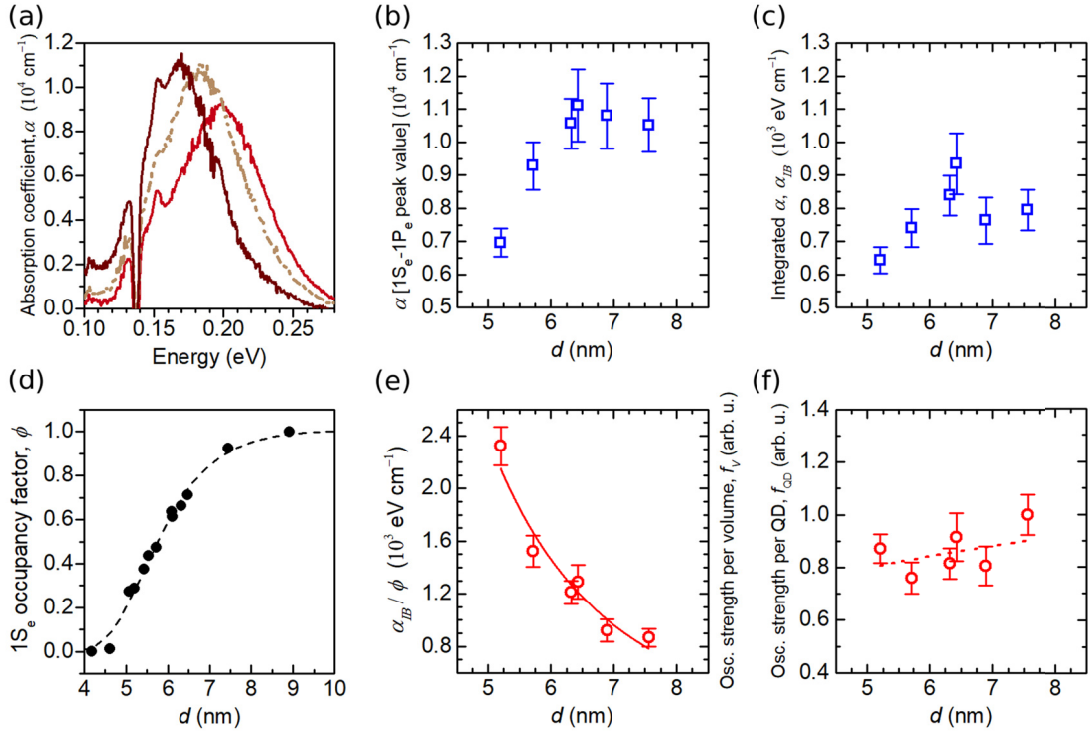

**Figure 4**. Absorption coefficients of the intraband transitions in heavily n-doped PbS QDs. (a) Absorption coefficient spectra measured in PbS QD thin films, for QDs of different sizes. (b) Peak value of the intraband absorption coefficient. Values are obtained by fitting the absorption spectra to Gaussians. The error bars account for the error propagation of the uncertainty in film thickness (see Supporting Table S1) and the absorption measurements (5%, estimated). (c) Integrated $1S_e$-$1P_e$ absorption coefficients. Values are obtained by integration of the fitted Gaussians. (d) Measured occupancy factor of $1S_e$. (e) Integrated intraband absorption coefficient divided by the occupancy factor of $1S_e$, and oscillator strength per volume of the intraband absorption. The line is a power law fit. (f) Oscillator strength per QD of the intraband absorption. The dotted line is a guide to the eye.

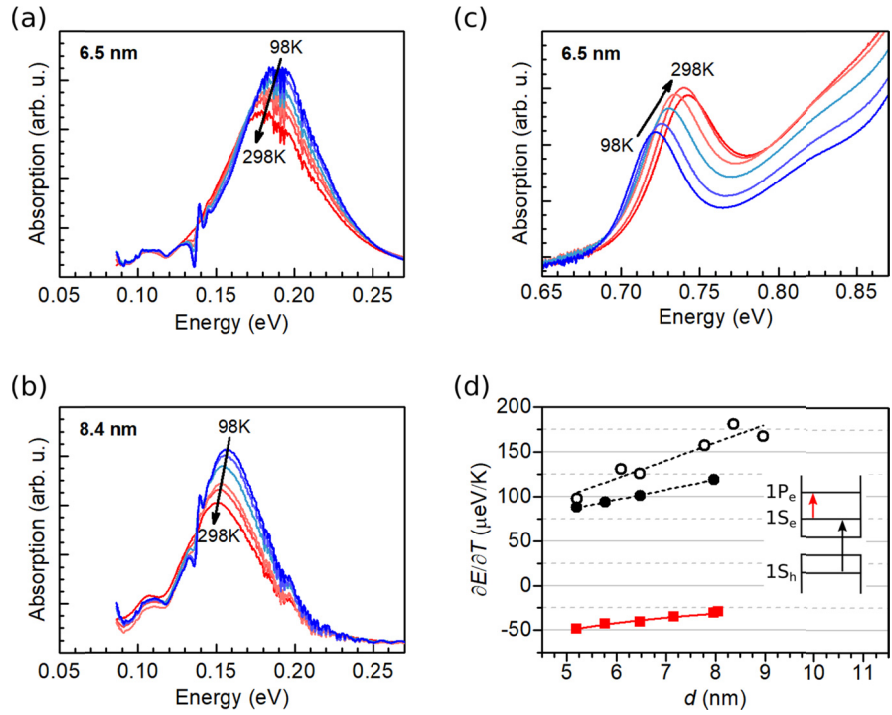

**Figure 5**. Intraband absorption spectrum at different temperatures of heavily n-doped PbS QDs with diameter of (a) 6.5 nm and (b) 8.4 nm. The two spikes at around 0.14 eV are due to the temperature variation of the $SiO_2$ absorption peak (see Methods). (c) Interband absorption spectrum at different temperatures of the same sample as in (a). (d) Temperature dependence of the interband (●) and intraband (■) transitions in heavily n-doped PbS QDs on silicon substrate. For comparison, the case of interband transitions in lightly n-doped QDs is also shown (○). Broken lines are guides to the eye. The solid red line is a fit to Eq. 8.

# Supporting Information

**Size and temperature dependent intraband optical studies of heavily n-doped PbS quantum dot solids**

Iñigo Ramiro[†,*], Biswajit Kundu[†], Mariona Dalmases[†], Onur Özdemir[†], María Pedrosa[†], and Gerasimos Konstantatos[†,‡,*]

[†]ICFO—Institut de Ciències Fotòniques, The Barcelona Institute of Science and Technology, Av. Carl Friedrich Gauss, 3, 08860 Castelldefels (Barcelona), Spain

[‡]ICREA—Institució Catalana de Recerca i Estudis Avançats, Passeig Lluís Companys 23, 08010 Barcelona, Spain

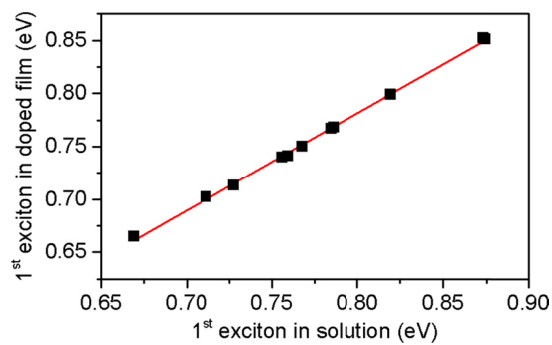

**Figure S1**. Energy shift of the 1st exciton in heavily n-doped PbS QD films with respect to the values in as synthesized PbS CQDs in solution. The red line is a linear fit to the measured data.

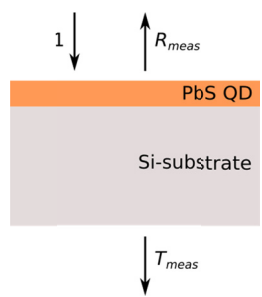

**Figure S2.** Structure of the samples used in absorption measurements. A thin film (100–130 nm) of heavily doped PbS QDs is coated on top of a semi-insulated Si substrate. 1 represents the 100% of the incoming light, $R_{meas}$ is the measured reflection of the sample, and $T_{meas}$ is the measured transmission of the sample.

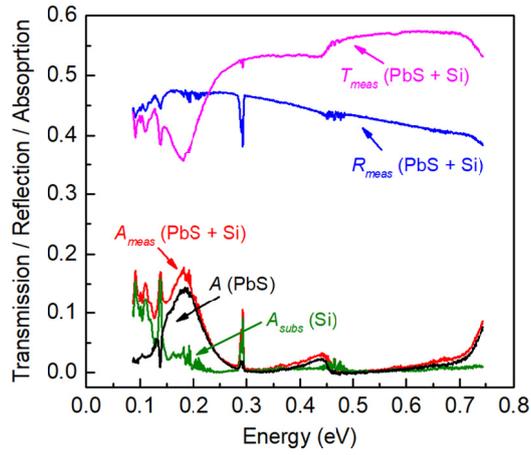

**Figure S3.** Example of the absorption measurement of a thin PbS layer in a sample as the one sketched in Figure S2. The absorption of the whole sample is obtained as: $A_{meas} = 1 - T_{meas} - R_{meas}$. The absorption of the PbS layer is obtained as: $A = A_{meas} - A_{subs}$.

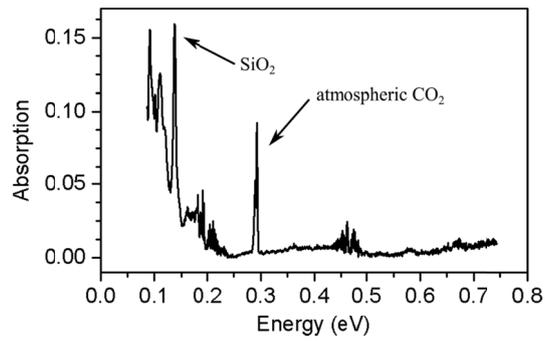

**Figure S4.** Typical absorption spectrum of the Si substrate used in this study.

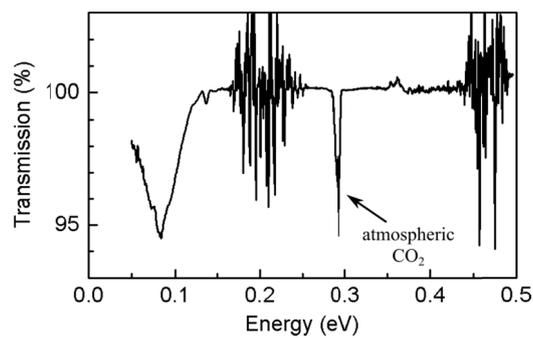

**Figure S5**. Absorption of a 20-nm thick layer of alumina deposited by ALD on a silicon substrate.

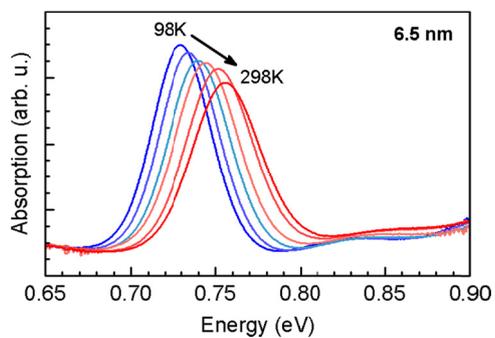

**Figure S6**. Temperature dependence of the interband absorption spectrum in lightly n-doped PbS QDs (shown for $d$ = 6.5 nm).

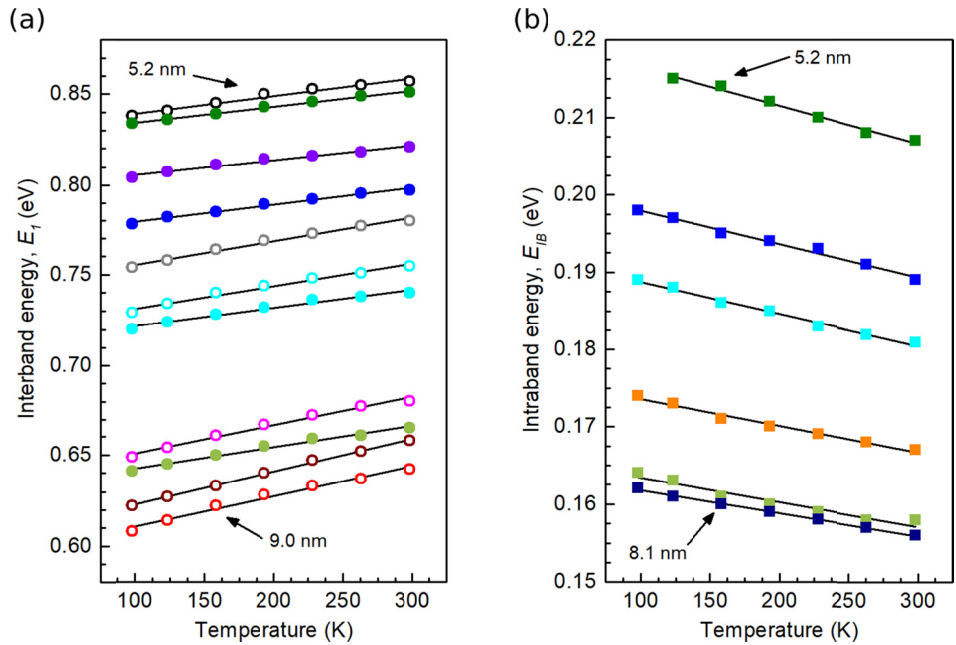

**Figure S7.** Interband (a) and intraband (b) energies in the 98–298 K temperature range for doped PbS QDs of different sizes. Full circles and squares represent interband and intraband energies, respectively, for heavily n-doped QDs. Open circles represent interband energies in lightly n-doped QDs. Same QD sizes in (a) and (b) are plotted with the same color. Solid lines are linear fits whose slopes represent the temperature dependence of the interband or intraband energies, $\partial E_1/\partial T$ or $E_{IB}/\partial T$.

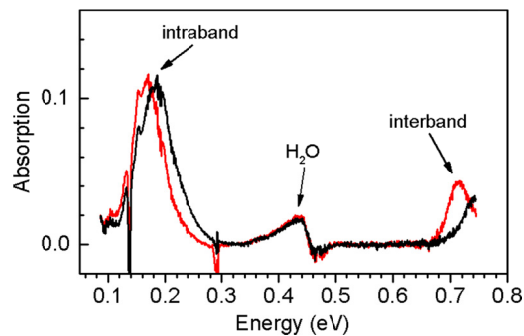

**Figure S8.** Absorption spectra of two films of heavily n-doped PbS QDs measured with an FTIR spectrometer. Both the 1st interband and the 1st intraband transitions can be identified. No contribution to the absorption from more energetic intraband transitions can be appreciated. An absorption band related to the presence of water is present in all samples. The water signature appears after deposition of alumina by ALD.

| $d$ (nm) | $t$ (nm) | $\phi$ | Peak $\alpha$ ($10^4$ cm$^{-1}$) | FWHM (meV) |
|---|---|---|---|---|
| 5.2 | 120 ± 2 | 0.28 | 0.69 | 87 |
| 5.7 | 125 ± 3 | 0.49 | 0.93 | 75 |
| 6.3 | 129 ± 2 | 0.69 | 1.06 | 75 |
| 6.4 | 100 ± 4 | 0.72 | 1.11 | 79 |
| 6.9 | 110 ± 5 | 0.83 | 1.08 | 66 |
| 7.6 | 121 ± 3 | 0.92 | 1.05 | 71 |

**Table S1**. Principal measured parameters of the samples analyzed in Figure 4 of the main text.